\documentclass{optica-article}

\journal{opticajournal} 

\articletype{Research Article}

\usepackage{lineno}
\usepackage{amsmath}
\usepackage{bm}
\usepackage{titlesec}
\usepackage{mathtools}
\usepackage{braket}
\usepackage{dutchcal}

\begin{document}

\title{Pulse characterization at the single-photon level through chronocyclic $Q$-function measurements}

\author{Abhinandan Bhattacharjee,\authormark{1,*} Patrick Folge,\authormark{1} Laura Serino,\authormark{1}  Jaroslav Řeháček,\authormark{2} Zdeněk Hradil,\authormark{2} Christine Silberhorn,\authormark{1} and Benjamin Brecht\authormark{1}}

\address{\authormark{1}Paderborn University, Integrated Quantum Optics, Institute for Photonic Quantum Systems (PhoQS), Warburger Straße 100, 33098 Paderborn, Germany\\
\authormark{2}Department of Optics, Palacký University, 17. listopadu 12, 77146 Olomouc, Czech Republic}

\email{\authormark{*}abhib@mail.uni-paderborn.de} 


\begin{abstract*} 
The characterization of the complex spectral amplitude that is, the spectrum and spectral phase, of single-photon-level light fields is a crucial capability for modern photonic quantum technologies. Since established pulse characterisation techniques are not applicable at low intensities, alternative approaches are required. Here, we demonstrate the retrieval of the complex spectral amplitude of single-photon-level light pulses through measuring their chronocyclic $Q-$function. Our approach draws inspiration from quantum state tomography by exploiting the analogy between quadrature phase space and time-frequency phase space. In the experiment, we perform time-frequency projections with a quantum pulse gate, which directly yield the chronocyclic $Q-$function. We evaluate the data with maximum likelihood estimation, which is the established technique for quantum state tomography. This yields not only an unambigious estimate of the complex spectral amplitude of the state under test that does not require any \textit{a priori}
information, but also allows for, in principle, estimating the spectral-temporal coherence properties of the state. Our method accurately recovers features such as jumps in the spectral phase and is resistant against regions with zero spectral intensity, which makes it immediately beneficial also for classical pulse characterization problems.  
\end{abstract*}

\section{Introduction}

The time-frequency (TF) degree of freedom of an ultrafast pulsed field at the single-photon level has gained significant attention due to its wide-ranging applications in quantum information science\cite{brechtprx2015, fabre2022pra, harder2017RSP, ansari2018optica}. 
These applications include spectroscopy \cite{mukamel2020JOPB}, metrology \cite{donohue2018prl, shah2021prapplied, fabre2021pra, triggiani2023prapplied}, and communication \cite{bouchard2022prx, wang2021ieee}. 
TF modes are compatible with integrated optics platforms and long-distance free-space propagation and offer a reliable foundation for scalable quantum information applications. 
To enhance the performance of these applications, it becomes crucial to accurately characterize a TF state at the single-photon level. 
Typically, a TF state is characterized by measuring its complex spectral amplitude of the electric field. 
Traditional techniques such as frequency-resolved optical gating (FROG) \cite{kane1993oplet, Trebino:93, paye1993optlet1, delong1994josab}, spectral phase interferometry for direct electric-field reconstruction (SPIDER) \cite{iaconis1998optlett, shuman1999optexp, dorrer2002josab} and their variants \cite{fittinghoff1996optlett, gallmann2001optlett, londero2003jmo, reid2000optlett, stibenz2005optlett, french2009optlet, baum2004optlett, birge2006optlett, Bourassin-Bouchet:13}\textemdash for a comprehensive review see \cite{monmayrant2010jopb}\textemdash are not well-suited for single-photon level measurements because these methods typically require high power input. The underlying reason is that these techniques require a time non-stationary element with response times of the order of the pulses under investigation \cite{walmsley1996josab}, which can be only achieved with nonlinear optics processes. 
There are, of course, alternative approaches that can work for low-level light pulses, such as chronocyclic tomography \cite{beck1993optlett, Dorrer:03} or two-photon spectral interferometry \cite{Thiel:20}. 
In addition, interferometric approaches based on electro-optic shearing interferometry (EOSI) have emerged as promising methods for measuring the complex spectral amplitude at the single-photon level \cite{davis2018pra, davis2018prl, thekkadath2022prl, kurzyna2022optexp}. 
However, these techniques introduce experimental complexities such as ensuring interferometric stability, employing spectral shearing or temporal phase manipulation using modulators, or performing spectrally resolved photon-counting measurements. 
These can limit the practical applicability of these techniques.

On the other hand, in the context of quantum state tomography (QST) in continuous variable (CV) quantum optics, quadrature phase space quasi-probability distributions such as Wigner and Husimi $Q-$functions are routinely used to describe a quantum state \cite{d1994pra, fiuravsek2015pra, landon2018prl, ahmed2021prl}. This approach has also been applied to other continuous degrees of freedom, including position-momentum \cite{waller2012phase} and time-frequency \cite{paye1992ieee}. In the TF phase space, these quasi-probability distributions are called chronocyclic (known as TF) Wigner \cite{paye1992ieee, beck1993optlett, monmayrant2010jopb} and $Q-$functions \cite{praxmeyer2007prl}. The chronocyclic Wigner function serves as a widely used theoretical tool for intuitively understanding the TF characteristics of single-photon states \cite{paye1992ieee, fabre2022pra, brecht2013pra}. However, it is important to note that the chronocyclic Wigner function can take on negative values.  As a result, direct measurement becomes impossible, necessitating the use of reconstruction algorithms \cite{beck1993optlett, dorrer2003optlett, paye1992ieee}. In contrast, the chronocyclic $Q-$function of a TF state represents its projection onto an informationally complete set of reference states and always yields positive values. This enables the direct measurement of the chronocyclic $Q-$function without the need for reconstruction algorithms and therefore offers a direct experimental access to the spectral-temporal characteristics of a TF state.

In this article, we propose and demonstrate an innovative approach for measuring chronocyclic $Q-$functions of pulsed single-photon-level TF states. We use an integrated waveguide device known as the quantum pulse gate (QPG) \cite{eckstein2011optexp, brecht2011njp}, a dispersion-engineered frequency-conversion process that accurately projects an input mode onto a user-chosen TF mode \cite{Ansaripra2017, allgaier2017QST, ansari2018prl, gil2021optical, ansari2020optexp}. 
Using the QPG, we project an input TF state onto Fourier-limited Gaussian TF modes with different temporal and spectral shifts, which play the role of the coherent state basis in CV QST. We count upconverted photons as a function of temporal and spectral shifts, which provides the values of the chronocyclic $Q-$function. To demonstrate the effectiveness of our technique, we measure chronocyclic $Q-$functions of single-photon level states having complex TF mode structures. These $Q-$functions contain the complete spectral information of the unknown TF state, which we extract by means of maximum-likelihood estimation (MLE) \cite{hradil1997pra}, a method that does not rely on algorithmic inversion or reconstruction and that unambiguously yields the spectral amplitude and phase of the measured TF state. 

\section{Concepts and Theory}

\subsection{Quadrature phase space Husimi $Q-$function and quantum state tomography}\label{Q-function}

In the quadrature phase space, the value of the Husimi $Q-$function of a quantum state that is described by a density matrix $\hat\rho$ is defined as the projection of that state onto a coherent state $|\alpha\rangle$:
\begin{equation}
    Q(\alpha) = \frac{1}{\pi}\langle\alpha|\hat\rho|\alpha\rangle.\label{Qfun-quad}
\end{equation}
Here, the complex amplitude $\alpha$ represents the displacement of the coherent state $|\alpha\rangle$ from the origin of the quadrature phase space. The Husimi $Q-$function $Q(\alpha)$ contains the complete information of the state $\hat\rho$ \cite{landon2018prl, ahmed2021prl}.  
However, using only Eq. (\ref{Qfun-quad}) we cannot reconstruct the quantum state $\hat\rho$ unambiguously {for pure states}. This is known as "phase retrieval problem" and it is described in the following manner.

Consider a pure state $|\psi\rangle$ with corresponding density matrix $\hat\rho=|\psi\rangle\langle\psi|$. The state $|\psi\rangle$ can be represented in the over-complete coherent state basis $\{|\beta\rangle\}$ as $|\psi\rangle=\int d^2\beta \psi(\beta)|\beta\rangle$, where $\psi(\beta)$ represents the complex amplitude that completely characterizes the state $|\psi\rangle$. The $Q-$function can then be expressed as
\begin{equation}
    Q(\alpha) = \frac{1}{\pi}\left\lvert\int d^2\beta\psi(\beta)\langle\alpha|\beta\rangle\right\rvert^2,\label{Qfun-quad1}
\end{equation}
where $\langle\alpha|\beta\rangle$ is a Gaussian kernel. The above equation can be rewritten as 
\begin{equation}
    \sqrt{Q(\alpha)}e^{i\theta(\alpha)} = \frac{1}{\sqrt\pi}\int d^2\beta\psi(\beta)\langle\alpha|\beta\rangle,\label{Qfun-quad2}
\end{equation}
where $\theta(\alpha)$ is an arbitrary function of $\alpha$. This phase function cannot be obtained from the experimental data or the measured $Q-$function and must be fixed by prior assumptions. {Consequently, Eq.~(\ref{Qfun-quad}) cannot be inverted unambiguously. To avoid this ambiguity, we reconstruct the quantum state in terms of a density matrix rather than a complex amplitude. We therefore apply QST for an umambiguous reconstruction that foregoes any \textit{a priori} knowledge.}

In general, QST yields a reconstructed density matrix $\hat\rho_r=\sum_n \lambda_n |\psi_n\rangle\langle\psi_n|$, which is a mixture of pure states $\{|\psi_n\rangle\}$ with corresponding weights $\{\lambda_n\}$. If this density matrix is reconstructed from a measured $Q-$function that describes a pure state, we expect a dominant contribution $\lambda_n$ from one of the $|\psi_n\rangle$ that represents the reconstructed pure state, while the remaining terms arise due to noise in the measurement. We note that MLE \cite{hradil1997pra, vrehavcek2007pra, banaszek1999pra}, which we also adapt in this work, is the state of the art methodology for reconstructing quantum states $\hat\rho$ from measured $Q-$functions. Here, however, we apply it to the problem of reconstructing the complex spectrum of an unknown TF state from a measured chronocyclic $Q-$ function.

\subsection{Time-frequency states}\label{TF-state}

In the following, we restrict our considerations to the case of single-photon-level TF states with perfect temporal coherence in a well-defined spatial and polarization mode. In the language of CV QST, this is analogous to a pure state. Note that this is a reasonable assumption when investigating ultrafast pulses that originate from a modelocked oscillator. The complex spectrum of such a TF state is then characterized by a single TF mode, which will be the dominant term in the reconstruction obtained from the chronocyclic $Q-$function, and which we will label as $f(\omega_{\rm in})=|f(\omega_{\rm in})|\exp[i\phi(\omega_{\rm in})]$. Here, $|f(\omega_{\rm in})|$ and $\phi(\omega_{\rm in})$ are the spectral amplitude and phase profiles, respectively.

\subsection{Analog of coherent state in time-frequency phase space}
To define a chronocyclic $Q-$function, we need an analog of a coherent state $|\alpha\rangle$ (see Eq. (\ref{Qfun-quad})) in the TF phase space. In quadrature phase space, a coherent state is displaced from the origin and has a two-dimensional Gaussian shape with equal uncertainties along both axes. We will now show that a Fourier-limited Gaussian TF mode of spectral width $\sigma_c/\sqrt{2}$ with spectral and temporal shifts  $(\omega^{(0)}_{\rm in},\tau_0)$ from the origin of the TF phase space is the TF analog of a coherent state in quadrature phase space. We write the corresponding complex spectral amplitude as
\begin{equation}
	\mathcal{E}_c(\omega_{\rm in},\omega,\tau;\sigma_c) = \exp\left[-\frac{(\omega_{\rm in}-\omega^{(0)}_{\rm in}-\omega)^2}{2\sigma_c^2}\right]\exp\left[i(\omega_{\rm in}-\omega^{(0)}_{\rm in})(\tau_0-\tau)\right],
\end{equation}
In this expression, $\omega^{(0)}_{\rm{in}}$ and $\tau_0$ define the origin of the TF phase space, and $\omega$ and $\tau$ are the spectral and temporal shifts, respectively. Without loss of generality, we set $\tau_0=0$. We find that spectral and temporal widths associated with  $\mathcal{E}_c(\omega_{\rm in},\omega,\tau;\sigma_c)$ are $\frac{\sigma_c}{\sqrt{2}}$ and $\frac{1}{\sqrt{2}\sigma_c}$ respectively. While the axes of the quadrature phase space are unitless, the axes of the TF phase space have units. Therefore, a rescaling of the frequency and time axes is a subtle necessity if one wants to draw an analogy between the two pictures. We choose a rescaling  $\xi=\frac{\omega}{\sigma_c}$, $t=\tau\sigma_c$, and $\xi_{\rm i}=\frac{\omega_{\rm i}}{\sigma_c}$ for which both widths equal $\frac{1}{\sqrt{2}}$. This emulates the symmetric uncertainties of a coherent state in quadrature phase space. In terms of these dimensionless, rescaled variables, the above TF mode becomes
\begin{equation}
	\mathcal{E}_c(\xi_{\rm in},\xi,t) = \exp\left[-\frac{(\xi^{(0)}_{\rm in}-\xi_{\rm in}-\xi)^2}{2}\right]\exp\left[-i(\xi^{(0)}_{\rm in}-\xi_{\rm in})t\right],
\end{equation}
which will serve as "coherent" state in dimensionless units. We will call this TF mode a \textit{TF coherent mode}. We note that any other rescaling yields unequal widths along the rescaled time and frequency axes, {resulting in an analog of a squeezed state in quadrature phase space. In this case, the reconstruction method would have to be adapted, which is why rescaling is crucially required.} 
In appendix~\ref{appen1}, we further show that the above $\mathcal{E}_c(\xi_{\rm in},\xi,t)$ can be expanded as a superposition of Hermite-Gaussian modes, analogous to the expansion of a coherent state $|\alpha\rangle$ as a superposition of photon-number Fock states $|n\rangle$.

\subsection{Chronocyclic $Q-$function}\label{Qfun}                                                                                                                                               
Now, we define the chronocyclic $Q-$function in dimensionless variables $Q(\xi,t)$ of a TF state with a complex spectrum $f({\xi_{\rm in}})$ as the projection of $f({\xi_{\rm in}})$ onto the coherent mode $\mathcal{E}_c(\xi_{\rm in}, \xi, t)$.
\begin{equation}
    Q(\xi, t) \propto \left\lvert\int f(\xi_{\rm in}) \mathcal{E}^{*}_c(\xi_{\rm in}, \xi, t) d\xi_{\rm in}\right\rvert^2.\label{chrono-Qfun}
\end{equation}
This definition of the chronocyclic $Q-$function is now fully analogous to that of the quadrature phase space Husimi $Q-$function in Eq.~(\ref{Qfun-quad}) and it contains the complete spectral information of the TF state.
\subsection{Reconstruction of complex spectral amplitude}\label{Qfun_reconstruction}
Because of the close analogy between quadrature phase space and TF phase space, the phase retrieval problem also applies to the chronocyclic $Q-$function. Consequently, an unambigious inversion of Eq. (\ref{chrono-Qfun}) for retrieving the complex spectrum $f(\xi_{\rm in})$ is not possible. Instead, we further lean on the analogy between quadrature and TF phase space and use MLE \cite{hradil1997pra, vrehavcek2007pra, banaszek1999pra} for reconstructing the complex spectrum of the TF state under investigation. 

The MLE reconstruction yields a two-point spectral correlation function $W_e(\omega'_{\rm in},\omega_{\rm in})$ that describes the spectral properties of the TF state and takes on the role of the density matrix $\hat\rho_r$ from Sec. \ref{Q-function}. We use the decomposition $W_e(\omega'_{\rm in},\omega_{\rm in})=\sum_n \lambda_n f^{*}_n(\omega'_{\rm in})f_n(\omega_{\rm in})$ to write the spectral two-point correlation function as a sum of complex spectra $\{f_n(\omega_{\rm in})\}$ with corresponding weights $\{\lambda_n\}$. For the case of TF states with perfect temporal coherence, the dominant contribution is the complex spectrum of the TF state, while the remaining contributions are again caused by measurement noise.

\begin{figure}[t!]
    \centering
    \includegraphics[scale=0.95]{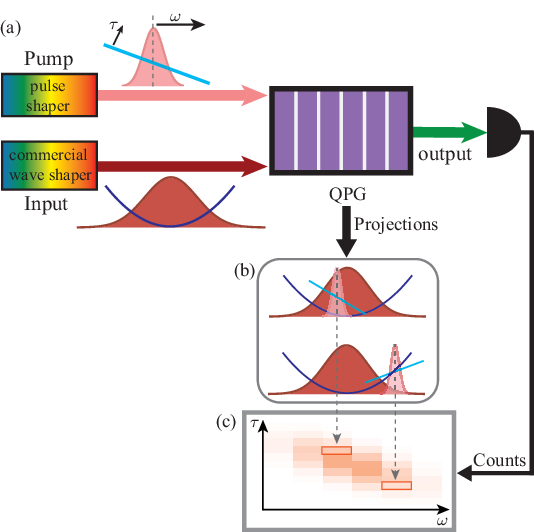}
    \caption{(a) Schematic of the scheme proposed for measuring chronocyclic $Q-$function. (b) Illustrating projection of the input on different spectrally and temporally shifted pump modes. (c) QPG output counts as a function of spectral and temporal shifts.}\label{fig1}
\end{figure} 
\subsection{Method for measuring chronocyclic $Q-$function}

In order to measure the chronocyclic $Q-$function $Q(\xi, t)$ of an arbitrary TF state, we need to perform projections onto TF coherent modes with spectral width $\sigma_c/\sqrt{2}$ that exhibit different spectral and temporal shifts ($\xi$ and $t$). 

For projective measurements in TF degrees of freedom, we utilize the quantum pulse gate (QPG), a device routinely used for this purpose \cite{brecht2014pra}. The QPG is a dispersion-engineered up-conversion process that enables the projection of an input into any desired TF mode. Our proposed scheme is described through the illustration in Fig.~\ref{fig1}. We employ two different pulse shapers to generate various input TF states $f(\xi_{\rm in})$ and TF coherent modes $\mathcal{E}_c(\xi_{\rm in},\xi,t)$ with different $\xi=\omega/\sigma_c$ and $t=\tau\cdot\sigma_c$ for pumping the QPG. An ideal QPG perfectly projects the input on the TF coherent mode, see Fig.~\ref{fig1}(b) and we routinely achieve projection fidelities greater than 96$\%$ \cite{Ansaripra2017, serino2023prx}. The total intensity of the QPG output $\eta$ provides the result of the projection and is written as
\begin{equation}
    \eta \propto \left\lvert \int f(\xi_{\rm in})\mathcal{E}^{*}_c(\xi_{\rm in}, \xi, t) d\xi_{\rm in}\right\rvert^2
\end{equation}
which is exactly the definition of $Q(\xi,t)$ given in Eq.~(\ref{chrono-Qfun}). This situation is illustrated in Fig.~\ref{fig1}(c).

\section{Experiment}

\begin{figure}[t!]
    \centering
    \includegraphics{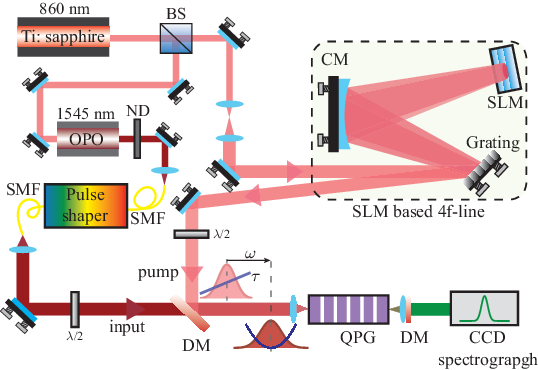}
    \caption{(a) Schematic of the experimental setup. BS: beam splitter, OPO: optical parametric oscillator, DM: dichroic mirror, SMF: single mode fiber, SLM: spatial light modulator, CM: cylindrical mirror.}\label{fig3}
\end{figure} 
Figure~\ref{fig3} shows the schematic of the experimental setup for measuring the chronocyclic $Q-$function of single-photon level input pulses. A titanium Sapphire pulsed laser of central wavelength 860 nm (349 THz) and repetition rate of 80 MHz is split using a beam-splitter. One portion is send to pump an OPO process in order to generate input pulses centered at 1545 nm. We use a commercial pulse shaper of resolution 1 GHz to shape these pulses with customized spectral amplitude and phase profiles. An ND filter attenuates the input to 2-3 photons per pulse. The remaining portion of 860 nm pulse is send to a home-built 4-f line based on
\begin{figure*}[t!]
    \centering
    \includegraphics[width=\linewidth]{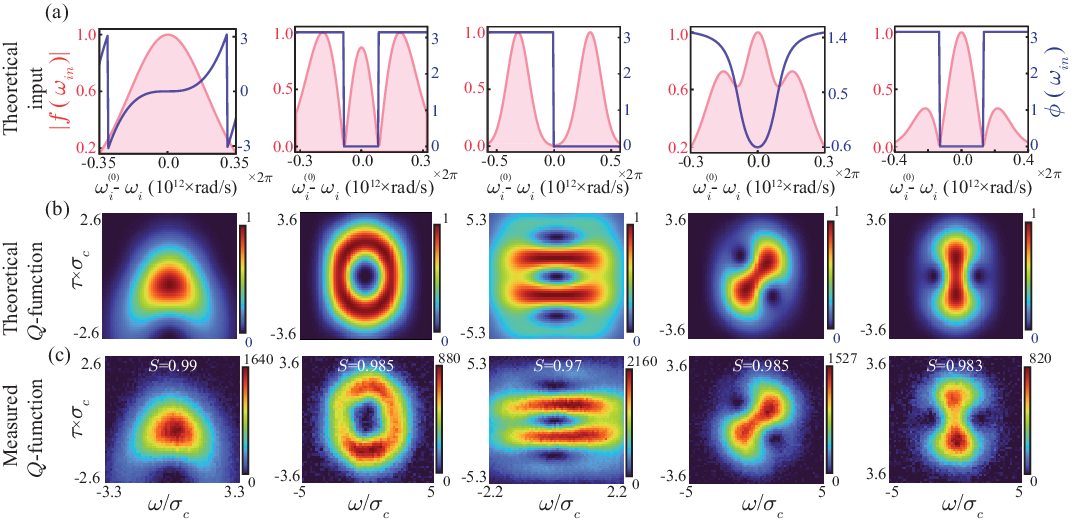}
    \caption{(a) Theoretical complex spectral amplitudes of different TF states. (b) and (c) Theoretical and measured chronocyclic $Q-$functions respectively, corresponding to different input modes.}\label{fig4}
\end{figure*} 
a spatial light modulator (SLM) with a resolution of 10 GHz, to spectrally shape it as a TF coherent mode and use it as the pump for the QPG. The spectral width $\sigma_c/{2\pi}$ varies in the range of $0.12-0.26$ THz. We also ensure that the pump does not have any second or higher order spectral phase components before it enters the QPG such that the pump is perfectly in the TF coherent mode. The half-waveplates ($\lambda/2$ plates) are placed in the path of both input and pump pulses to align that their polarization states appropriately for the QPG process. 

We use a $40$ mm long Ti-indiffused LiNbO$_3$ QPG waveguide with a poling period of $4.32$ $\mu$m and operate at 433 K to ensure group velocity matching between both pulses. The phase-matching width of the QPG is $0.06\times2\pi$ rad/s. The waveguide is designed to couple the fundamental spatial mode for the input wavelength, and we make sure that the pump wavelength also couples into the waveguide in the fundamental spatial mode. The couple efficiencies for both the input and pump pulses are approximately 60\%. 

The output of the QPG, centered at 552 nm (543 THz) is separated from the residual pump and input fields using a dichroic mirror (DM). We measure it using a commercial single-photon sensitive CCD spectrograph $($Andor Shamrock 500i$)$ with a resolution of 30 GHz. The integration time varies between 1.4 and 4 seconds. As previously mentioned the QPG has a finite phase-matching width and in order to ensure high quality projections, we perform wavelength filtering on the measured QPG output \cite{santandrea2019njp,serino2023prx}. 

Figure~\ref{fig4}(a) shows the theoretical complex spectral amplitude profiles of different single-photon level input pulses. These include features such as phase jumps and regions of zero spectral intensity (c.f. the middle column) that typically pose a challenge for pulse characterization schemes based on algorithmic reconstruction. For each input TF state, we vary the spectral and temporal shifts ($\xi$ and $t$) of the pump using the SLM-based $4f$-line pulse shaper. We obtain the $Q-$function by measuring the total intensity of the QPG output as a function of these shifts. We also record the background noise profile, which we subtract from the measured QPG output to obtain the experimental chronocyclic $Q-$function. We note that the CCD spectrograph is used as a photon counter; we record integrated counts over the whole spectrum, hence do not employ spectrally resolved measurements. The use of the spectrograph is for experimental convenience and it could readily be replaced with single-photon detectors.

\section{Results}
Figures~\ref{fig4}(b) and ~\ref{fig4}(c) show the theoretical and measured chronocyclic $Q-$functions, respectively, for the different single-photon-level input TF states. We scale the maximum of the theoretical plots to one and the measured plots are scaled with the QPG output counts. For a first quantification of the match between theory and experiment, we use a quantity known as the similarity $S$ (see Appendix~\ref{appen2}). The similarity $S$ lies between $0$ and $1$, $S=0$ indicates no match and $S=1$ indicates a perfect match. In Fig.~\ref{fig4}(c), we show $S$ for each measured $Q-$function. We find a very good quantitative agreement between theoretical and measured results with similarities above $97\%$. The slight mismatch between theory and measurements can be attributed to the following reasons. The intrinsic spectral envelopes of both input and pump TF mode can introduce imperfections in shaping their respective spectral amplitude and phase profiles. Any uncompensated spectral phase component in the pump can cause imperfect projections. Moreover, a finite phase-matching width (60 GHz) of the QPG as compared to that of the coherent mode (120-260 GHz) can also cause imperfect projections. One can address these issues by increasing the shaping resolution of input and pump TF states and using a QPG waveguide with a much narrower phase-matching width.
\begin{figure*}[t!]
    \centering
    \includegraphics[width=\linewidth]{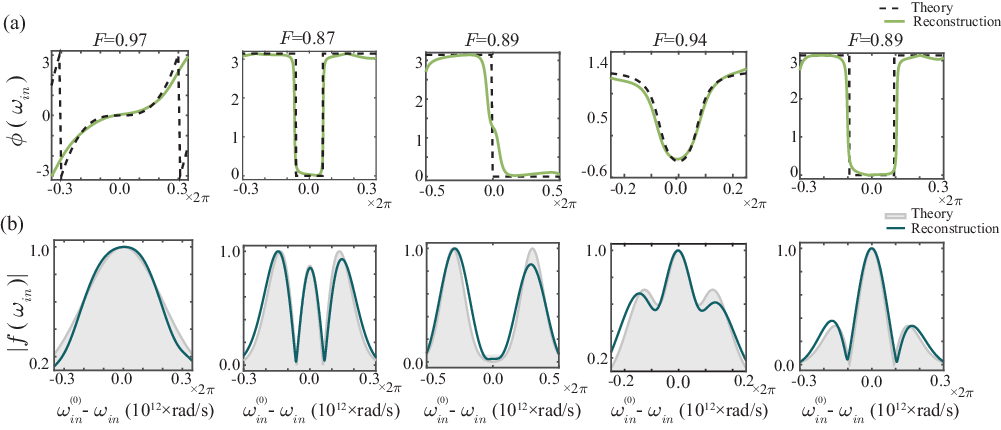}
    \caption{ (a) Theoretical and reconstructed spectral phase profiles. (b) Theoretical and reconstructed spectral amplitude profiles.}\label{fig5}
\end{figure*} 

We now retrieve the spectral phase and amplitude profiles from the measured chronocyclic $Q-$function using MLE. The estimated TF states expressed through two-point spectral correlation functions $W_e(\omega'_{\rm in},\omega_{\rm in})$ are shown in appendix~\ref{appen2}. In order to quantify the quality of our reconstruction, we evaluate the fidelity ${F}$ between the estimated and the input TF state (see appendix~\ref{appen2}). As we have mentioned earlier the $W_e(\omega'_{\rm in},\omega_{\rm in})$ is a sum of complex spectral amplitudes $\{f_n(\omega_{\rm in})\}$ with weights ${\lambda_n}$. In order to extract the spectral amplitude and phase profile from $W_e(\omega'_{\rm in},\omega_{\rm in})$, we choose $f_n(\omega_{\rm in})$ corresponding to the dominant $\lambda_n$. The obtained $f_n(\omega_{\rm in})$ is the reconstructed complex spectral amplitude.

The reconstructed profiles are shown in Fig.~\ref{fig5} alongside their theoretical shapes. We scale both theoretical and reconstructed spectral amplitude profiles such that their maximum value is 1. 

The deviation between theoretical and reconstructed results can arise due to imperfections involved in the measurement of the chronocyclic $Q-$functions as detailed above. Similar issues have also been reported in CV state tomography algorithms based on Husimi $Q-$function measurements \cite{landon2018prl}.

\section{Summary}
In summary, we have proposed and demonstrated an experimental scheme for measuring the chronocyclic $Q-$function of pulsed single photon level TF states using a QPG. We have measured chronocyclic $Q-$functions for TF states having various spectral shapes at the single-photon level. We further showcase the high-quality reconstruction of spectral amplitude and phase profiles from the measured $Q-$functions, even for complex pulse shapes that include discontinuous phase profiles. Our approach draws inspiration from CV quantum optics by employing MLE for TF state reconstruction. This facilitates the reconstruction of complex spectra without \textit{a priori} assumptions and without resorting to inversion {effectively addressing the "phase retrieval problem" through QST.}  

Compared to existing EOSI-based interferometric approaches, our technique offers some advantages. The use of the QPG, which is an integrated waveguide device, makes our scheme, highly compatible with large scale quantum information applications. The QPG eliminates the need for phase stabilization, a common challenge in interferometric schemes. Our method can potentially be applied to wavelengths where single-photon detectors are unavailable, such as in the characterization of mid-infrared (MIR) TF states. 

Furthermore, our approach is fully compatible with pulses exhibiting arbitrary temporal coherence, which translates to states of arbitrary spectral purity in quantum optics. This is a significant benefit over existing pulse characterization schemes that require coherence. We expect that many quantum information that require accurate characterization of single-photon-level TF states will benefit from our approoach. 

\section{Acknowledgement}

\begin{backmatter}
\bmsection{Funding}
This project has received funding from the European Union’s Horizon Europe research and innovation programme under grant agreement No 899587 (STORMYTUNE). 

\bmsection{Acknowledgments}
We thank Prof. Dr. Jan Sperling for helpful and inspiring discussions.

\bmsection{Disclosures}
The authors declare no conflicts of interest.

\bmsection{Data availability} Data underlying the results presented in this paper are not publicly available at this time but may be obtained from the authors upon reasonable request.
\end{backmatter}

\bibliography{ref}

\begin{thebibliography}{10}
\newcommand{\enquote}[1]{``#1''}

\bibitem{brechtprx2015}
B.~Brecht, D.~V. Reddy, C.~Silberhorn, and M.~G. Raymer, \enquote{Photon
  temporal modes: A complete framework for quantum information science,}
  {\protect\JournalTitle{Phys. Rev. X}} \textbf{5}, 041017 (2015).

\bibitem{fabre2022pra}
N.~Fabre, A.~Keller, and P.~Milman, \enquote{Time and frequency as quantum
  continuous variables,} {\protect\JournalTitle{Physical Review A}}
  \textbf{105}, 052429 (2022).

\bibitem{harder2017RSP}
G.~Harder, V.~Ansari, T.~Bartley, \emph{et~al.}, \enquote{Harnessing temporal
  modes for multi-photon quantum information processing based on integrated
  optics,} {\protect\JournalTitle{Philosophical Transactions of the Royal
  Society A: Mathematical, Physical and Engineering Sciences}} \textbf{375},
  20160244 (2017).

\bibitem{ansari2018optica}
V.~Ansari, J.~M. Donohue, B.~Brecht, and C.~Silberhorn, \enquote{Tailoring
  nonlinear processes for quantum optics with pulsed temporal-mode encodings,}
  {\protect\JournalTitle{Optica}} \textbf{5}, 534--550 (2018).

\bibitem{mukamel2020JOPB}
S.~Mukamel, M.~Freyberger, W.~Schleich, \emph{et~al.}, \enquote{Roadmap on
  quantum light spectroscopy,} {\protect\JournalTitle{Journal of physics B:
  Atomic, molecular and optical physics}} \textbf{53}, 072002 (2020).

\bibitem{donohue2018prl}
J.~M. Donohue, V.~Ansari, J.~{\v{R}}eh{\'a}{\v{c}}ek, \emph{et~al.},
  \enquote{Quantum-limited time-frequency estimation through mode-selective
  photon measurement,} {\protect\JournalTitle{Physical review letters}}
  \textbf{121}, 090501 (2018).

\bibitem{shah2021prapplied}
M.~Shah and L.~Fan, \enquote{Frequency superresolution with spectrotemporal
  shaping of photons,} {\protect\JournalTitle{Physical Review Applied}}
  \textbf{15}, 034071 (2021).

\bibitem{fabre2021pra}
N.~Fabre and S.~Felicetti, \enquote{Parameter estimation of time and frequency
  shifts with generalized hong-ou-mandel interferometry,}
  {\protect\JournalTitle{Physical Review A}} \textbf{104}, 022208 (2021).

\bibitem{triggiani2023prapplied}
D.~Triggiani, G.~Psaroudis, and V.~Tamma, \enquote{Ultimate quantum sensitivity
  in the estimation of the delay between two interfering photons through
  frequency-resolving sampling,} {\protect\JournalTitle{Physical Review
  Applied}} \textbf{19}, 044068 (2023).

\bibitem{bouchard2022prx}
F.~Bouchard, D.~England, P.~J. Bustard, \emph{et~al.}, \enquote{Quantum
  communication with ultrafast time-bin qubits,} {\protect\JournalTitle{PRX
  Quantum}} \textbf{3}, 010332 (2022).

\bibitem{wang2021ieee}
Z.~Wang, R.~Malaney, and R.~Aguinaldo, \enquote{Temporal modes of light in
  satellite-to-earth quantum communications,} {\protect\JournalTitle{IEEE
  Communications Letters}} \textbf{26}, 311--315 (2021).

\bibitem{kane1993oplet}
D.~J. Kane and R.~Trebino, \enquote{Single-shot measurement of the intensity
  and phase of an arbitrary ultrashort pulse by using frequency-resolved
  optical gating,} {\protect\JournalTitle{Optics letters}} \textbf{18},
  823--825 (1993).

\bibitem{Trebino:93}
R.~Trebino and D.~J. Kane, \enquote{Using phase retrieval to measure the
  intensity and phase of ultrashort pulses: frequency-resolved optical gating,}
  {\protect\JournalTitle{J. Opt. Soc. Am. A}} \textbf{10}, 1101--1111 (1993).

\bibitem{paye1993optlet1}
J.~Paye, M.~Ramaswamy, J.~G. Fujimoto, and E.~P. Ippen, \enquote{Measurement of
  the amplitude and phase of ultrashort light pulses from spectrally resolved
  autocorrelation,} {\protect\JournalTitle{Optics letters}} \textbf{18},
  1946--1948 (1993).

\bibitem{delong1994josab}
K.~DeLong, R.~Trebino, J.~Hunter, and W.~White, \enquote{Frequency-resolved
  optical gating with the use of second-harmonic generation,}
  {\protect\JournalTitle{JOSA B}} \textbf{11}, 2206--2215 (1994).

\bibitem{iaconis1998optlett}
C.~Iaconis and I.~A. Walmsley, \enquote{Spectral phase interferometry for
  direct electric-field reconstruction of ultrashort optical pulses,}
  {\protect\JournalTitle{Optics letters}} \textbf{23}, 792--794 (1998).

\bibitem{shuman1999optexp}
T.~M. Shuman, M.~E. Anderson, J.~Bromage, \emph{et~al.}, \enquote{Real-time
  spider: ultrashort pulse characterization at 20 hz,}
  {\protect\JournalTitle{Optics Express}} \textbf{5}, 134--143 (1999).

\bibitem{dorrer2002josab}
C.~Dorrer and I.~A. Walmsley, \enquote{Accuracy criterion for ultrashort pulse
  characterization techniques: application to spectral phase interferometry for
  direct electric field reconstruction,} {\protect\JournalTitle{JOSA B}}
  \textbf{19}, 1019--1029 (2002).

\bibitem{fittinghoff1996optlett}
D.~N. Fittinghoff, J.~L. Bowie, J.~N. Sweetser, \emph{et~al.},
  \enquote{Measurement of the intensity and phase of ultraweak, ultrashort
  laser pulses,} {\protect\JournalTitle{Optics letters}} \textbf{21}, 884--886
  (1996).

\bibitem{gallmann2001optlett}
L.~Gallmann, G.~Steinmeyer, D.~Sutter, \emph{et~al.}, \enquote{Spatially
  resolved amplitude and phase characterization of femtosecond optical pulses,}
  {\protect\JournalTitle{Optics Letters}} \textbf{26}, 96--98 (2001).

\bibitem{londero2003jmo}
P.~Londero, M.~E. Anderson, C.~Radzewicz, \emph{et~al.}, \enquote{Measuring
  ultrafast pulses in the near-ultraviolet using spectral phase interferometry
  for direct electric field reconstruction,} {\protect\JournalTitle{Journal of
  Modern Optics}} \textbf{50}, 179--184 (2003).

\bibitem{reid2000optlett}
D.~Reid, P.~Loza-Alvarez, C.~Brown, \emph{et~al.}, \enquote{Amplitude and phase
  measurement of mid-infrared femtosecond pulses by using cross-correlation
  frequency-resolved optical gating,} {\protect\JournalTitle{Optics letters}}
  \textbf{25}, 1478--1480 (2000).

\bibitem{stibenz2005optlett}
G.~Stibenz and G.~Steinmeyer, \enquote{Interferometric frequency-resolved
  optical gating,} {\protect\JournalTitle{Optics express}} \textbf{13},
  2617--2626 (2005).

\bibitem{french2009optlet}
D.~French, C.~Dorrer, and I.~Jovanovic, \enquote{Two-beam spider for dual-pulse
  single-shot characterization,} {\protect\JournalTitle{Optics letters}}
  \textbf{34}, 3415--3417 (2009).

\bibitem{baum2004optlett}
P.~Baum, S.~Lochbrunner, and E.~Riedle, \enquote{Zero-additional-phase spider:
  full characterization of visible and sub-20-fs ultraviolet pulses,}
  {\protect\JournalTitle{Optics letters}} \textbf{29}, 210--212 (2004).

\bibitem{birge2006optlett}
J.~R. Birge, R.~Ell, and F.~X. K{\"a}rtner, \enquote{Two-dimensional spectral
  shearing interferometry for few-cycle pulse characterization,}
  {\protect\JournalTitle{Optics letters}} \textbf{31}, 2063--2065 (2006).

\bibitem{Bourassin-Bouchet:13}
C.~Bourassin-Bouchet, M.~M. Mang, I.~Gianani, and I.~A. Walmsley,
  \enquote{Mutual interferometric characterization of a pair of independent
  electric fields,} {\protect\JournalTitle{Opt. Lett.}} \textbf{38}, 5299--5302
  (2013).

\bibitem{monmayrant2010jopb}
A.~Monmayrant, S.~Weber, and B.~Chatel, \enquote{A newcomer's guide to
  ultrashort pulse shaping and characterization,}
  {\protect\JournalTitle{Journal of Physics B: Atomic, Molecular and Optical
  Physics}} \textbf{43}, 103001 (2010).

\bibitem{walmsley1996josab}
I.~A. Walmsley and V.~Wong, \enquote{Characterization of the electric field of
  ultrashort optical pulses,} {\protect\JournalTitle{JOSA B}} \textbf{13},
  2453--2463 (1996).

\bibitem{beck1993optlett}
M.~Beck, M.~Raymer, I.~Walmsley, and V.~Wong, \enquote{Chronocyclic tomography
  for measuring the amplitude and phase structure of optical pulses,}
  {\protect\JournalTitle{Optics letters}} \textbf{18}, 2041--2043 (1993).

\bibitem{Dorrer:03}
C.~Dorrer and I.~Kang, \enquote{Complete temporal characterization of short
  optical pulses by simplified chronocyclic tomography,}
  {\protect\JournalTitle{Opt. Lett.}} \textbf{28}, 1481--1483 (2003).

\bibitem{Thiel:20}
V.~Thiel, A.~O.~C. Davis, K.~Sun, \emph{et~al.}, \enquote{Single-photon
  characterization by two-photon spectral interferometry,}
  {\protect\JournalTitle{Opt. Express}} \textbf{28}, 19315--19324 (2020).

\bibitem{davis2018pra}
A.~O. Davis, V.~Thiel, M.~Karpi{\'n}ski, and B.~J. Smith, \enquote{Experimental
  single-photon pulse characterization by electro-optic shearing
  interferometry,} {\protect\JournalTitle{Physical Review A}} \textbf{98},
  023840 (2018).

\bibitem{davis2018prl}
A.~O. Davis, V.~Thiel, M.~Karpi{\'n}ski, and B.~J. Smith, \enquote{Measuring
  the single-photon temporal-spectral wave function,}
  {\protect\JournalTitle{Physical review letters}} \textbf{121}, 083602 (2018).

\bibitem{thekkadath2022prl}
G.~Thekkadath, B.~Bell, R.~Patel, \emph{et~al.}, \enquote{Measuring the joint
  spectral mode of photon pairs using intensity interferometry,}
  {\protect\JournalTitle{Physical Review Letters}} \textbf{128}, 023601 (2022).

\bibitem{kurzyna2022optexp}
S.~Kurzyna, M.~Jastrz{\k{e}}bski, N.~Fabre, \emph{et~al.}, \enquote{Variable
  electro-optic shearing interferometry for ultrafast single-photon-level pulse
  characterization,} {\protect\JournalTitle{Optics Express}} \textbf{30},
  39826--39839 (2022).

\bibitem{d1994pra}
G.~d~Ariano, C.~Macchiavello, and M.~Paris, \enquote{Detection of the density
  matrix through optical homodyne tomography without filtered back projection,}
  {\protect\JournalTitle{Physical Review A}} \textbf{50}, 4298 (1994).

\bibitem{fiuravsek2015pra}
J.~Fiur{\'a}{\v{s}}ek, \enquote{Continuous-variable quantum process tomography
  with squeezed-state probes,} {\protect\JournalTitle{Physical Review A}}
  \textbf{92}, 022101 (2015).

\bibitem{landon2018prl}
O.~Landon-Cardinal, L.~C. Govia, and A.~A. Clerk, \enquote{Quantitative
  tomography for continuous variable quantum systems,}
  {\protect\JournalTitle{Physical review letters}} \textbf{120}, 090501 (2018).

\bibitem{ahmed2021prl}
S.~Ahmed, C.~S. Mu{\~n}oz, F.~Nori, and A.~F. Kockum, \enquote{Quantum state
  tomography with conditional generative adversarial networks,}
  {\protect\JournalTitle{Physical Review Letters}} \textbf{127}, 140502 (2021).

\bibitem{waller2012phase}
L.~Waller, G.~Situ, and J.~W. Fleischer, \enquote{Phase-space measurement and
  coherence synthesis of optical beams,} {\protect\JournalTitle{Nature
  Photonics}} \textbf{6}, 474--479 (2012).

\bibitem{paye1992ieee}
J.~Paye, \enquote{The chronocyclic representation of ultrashort light pulses,}
  {\protect\JournalTitle{IEEE Journal of Quantum Electronics}} \textbf{28},
  2262--2273 (1992).

\bibitem{praxmeyer2007prl}
L.~Praxmeyer, P.~Wasylczyk, C.~Radzewicz, and K.~W{\'o}dkiewicz,
  \enquote{Time-frequency domain analogues of phase space sub-planck
  structures,} {\protect\JournalTitle{Physical review letters}} \textbf{98},
  063901 (2007).

\bibitem{brecht2013pra}
B.~Brecht and C.~Silberhorn, \enquote{Characterizing entanglement in pulsed
  parametric down-conversion using chronocyclic wigner functions,}
  {\protect\JournalTitle{Physical Review A}} \textbf{87}, 053810 (2013).

\bibitem{dorrer2003optlett}
C.~Dorrer and I.~Kang, \enquote{Complete temporal characterization of short
  optical pulses by simplified chronocyclic tomography,}
  {\protect\JournalTitle{Optics letters}} \textbf{28}, 1481--1483 (2003).

\bibitem{eckstein2011optexp}
A.~Eckstein, B.~Brecht, and C.~Silberhorn, \enquote{A quantum pulse gate based
  on spectrally engineered sum frequency generation,}
  {\protect\JournalTitle{Optics express}} \textbf{19}, 13770--13778 (2011).

\bibitem{brecht2011njp}
B.~Brecht, A.~Eckstein, A.~Christ, \emph{et~al.}, \enquote{From quantum pulse
  gate to quantum pulse shaper engineered frequency conversion in nonlinear
  optical waveguides,} {\protect\JournalTitle{New Journal of Physics}}
  \textbf{13}, 065029 (2011).

\bibitem{Ansaripra2017}
V.~Ansari, G.~Harder, M.~Allgaier, \emph{et~al.}, \enquote{Temporal-mode
  measurement tomography of a quantum pulse gate,} {\protect\JournalTitle{Phys.
  Rev. A}} \textbf{96}, 063817 (2017).

\bibitem{allgaier2017QST}
M.~Allgaier, G.~Vigh, V.~Ansari, \emph{et~al.}, \enquote{Fast time-domain
  measurements on telecom single photons,} {\protect\JournalTitle{Quantum
  Science and Technology}} \textbf{2}, 034012 (2017).

\bibitem{ansari2018prl}
V.~Ansari, J.~M. Donohue, M.~Allgaier, \emph{et~al.}, \enquote{Tomography and
  purification of the temporal-mode structure of quantum light,}
  {\protect\JournalTitle{Physical review letters}} \textbf{120}, 213601 (2018).

\bibitem{gil2021optical}
J.~Gil-Lopez, Y.~S. Teo, S.~De, \emph{et~al.}, \enquote{Universal compressive
  tomography in the time-frequency domain,} {\protect\JournalTitle{Optica}}
  \textbf{8}, 1296--1305 (2021).

\bibitem{ansari2020optexp}
V.~Ansari, J.~M. Donohue, B.~Brecht, and C.~Silberhorn, \enquote{Remotely
  projecting states of photonic temporal modes,} {\protect\JournalTitle{Optics
  Express}} \textbf{28}, 28295--28305 (2020).

\bibitem{hradil1997pra}
Z.~Hradil, \enquote{Quantum-state estimation,} {\protect\JournalTitle{Physical
  Review A}} \textbf{55}, R1561 (1997).

\bibitem{vrehavcek2007pra}
J.~{\v{R}}eh{\'a}{\v{c}}ek, Z.~Hradil, E.~Knill, and A.~Lvovsky,
  \enquote{Diluted maximum-likelihood algorithm for quantum tomography,}
  {\protect\JournalTitle{Physical Review A}} \textbf{75}, 042108 (2007).

\bibitem{banaszek1999pra}
K.~Banaszek, G.~Dariano, M.~Paris, and M.~Sacchi, \enquote{Maximum-likelihood
  estimation of the density matrix,} {\protect\JournalTitle{Physical Review A}}
  \textbf{61}, 010304 (1999).

\bibitem{brecht2014pra}
B.~Brecht, A.~Eckstein, R.~Ricken, \emph{et~al.}, \enquote{Demonstration of
  coherent time-frequency schmidt mode selection using dispersion-engineered
  frequency conversion,} {\protect\JournalTitle{Physical Review A}}
  \textbf{90}, 030302 (2014).

\bibitem{serino2023prx}
L.~Serino, J.~Gil-Lopez, M.~Stefszky, \emph{et~al.}, \enquote{Realization of a
  multi-output quantum pulse gate for decoding high-dimensional temporal modes
  of single-photon states,} {\protect\JournalTitle{PRX quantum}} \textbf{4},
  020306 (2023).

\bibitem{santandrea2019njp}
M.~Santandrea, M.~Stefszky, V.~Ansari, and C.~Silberhorn, \enquote{Fabrication
  limits of waveguides in nonlinear crystals and their impact on quantum optics
  applications,} {\protect\JournalTitle{New Journal of Physics}} \textbf{21},
  033038 (2019).

\end{thebibliography}

\appendix
\section{Analog between TF coherent mode and coherent state}\label{appen1}
To further establish the analog between TF coherent mode and coherent state, we show that the TF coherent mode can be expressed as the superposition of Hermite functions in the same manner as the coherent state as the superposition of fock states. The fock states are represented as orthogonal Hermite functions in the quadrature phase space. The chronocyclic $Q-$function can be expressed as 
%
\begin{equation}
	Q(\xi,t) \propto \left\lvert\int f(\xi_{\rm in})\mathcal{E}^{*}_c(\xi_{\rm in},\xi,t) d\xi_{\rm in}\right\rvert^2,\label{chrono-Qfun-app}
\end{equation}
where $\xi$ and $t$ representing spectral and temporal shifts respectively and $\mathcal{E}_c(\xi_{\rm in},\xi,t)=e^{-\frac{1}{2}(\xi^{(0)}_{\rm in}-\xi_{\rm in}-\xi)^2}e^{-i\xi_{\rm in}t}$. 
For the sake of convenience, we take $\tilde\xi_{\rm in}=\xi^{(0)}_{\rm in}-\xi_{\rm in}$. Using the following identity
\begin{align}
	-(\tilde\xi_{\rm in}-\xi)^2+2i\xi_{\rm in}t= -(\tilde\xi_{\rm in}-\xi-it)^2-t^2+2it\xi \label{identity}
\end{align}
we have 
\begin{equation}
	Q(\xi,t) \propto \exp\left(-\frac{t^2}{2}\right)\left\lvert\int f(\tilde\xi_{\rm in}) \exp\left[-\frac{(\tilde\xi_{\rm in}-\xi-it)^2}{2}\right] d\tilde\xi_{\rm in}\right\rvert^2.\label{chrono-Qfun-app-re1}
\end{equation}
Since
\begin{equation}
	\exp\left[-(x-y)^2\right] = \exp\left[-x^2\right]\sum_{m} \frac{1}{m!} H_m(x)y^m.\label{hermite-re1}
\end{equation}
we have
\begin{equation}
	Q(\xi,t) \propto \left\lvert \int f(\tilde\xi_{\rm in})\left[\sum_m \frac{1}{m!}\exp\left(-\frac{t^2}{2}\right)\exp\left(-\frac{\tilde\xi_{\rm in}^2}{2}\right)H_m\left(\frac{\tilde\xi_{\rm in}}{\sqrt{2}}\right)\beta^m\right] d\tilde\xi_{\rm in}\right\rvert^2.\label{chrono-Qfun-app-re2}
\end{equation}
where complex amplitude is $\beta = \frac{\xi+it}{\sqrt{2}}$. 
The above expansion is similar to the well-known expansion of a coherent state into the Fock state basis. This expansion is possible because we choose a TF coherent mode with equal variance along the rescaled $\xi$ and $t$ axes. 
\section{Estimated two-point cross spectral density function}\label{appen2}
\begin{figure*}[t!]
	\centering
	\includegraphics[width=\linewidth]{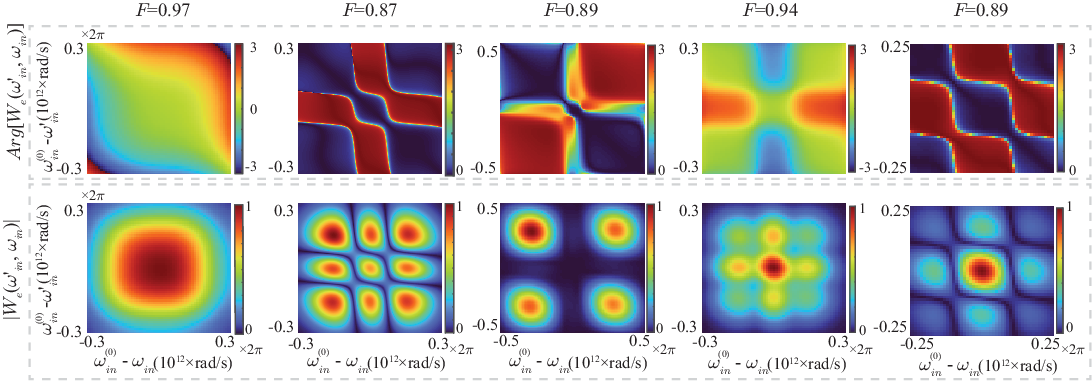}
	\caption{Estimated phase profile $Arg[W_e(\omega'_{in},\omega_{in})]$ and amplitude profile $|W_e(\omega'_{in},\omega_{in})|$ of two-point spectral correlation function respectively, for different input TF states.}\label{fig6}
\end{figure*} 
The MLE algorithm the TF state from the measured $Q-$function as a mixture of complex spectral amplitudes $f_n(\omega_{\rm in})$ with weightages $\lambda_n$ and it is expressed through the the two-point spectral correlation function $W(\omega'_{\rm in},\omega_{\rm in})$ as
\begin{equation}
	W_e(\omega'_{\rm in},\omega_{\rm in})=\sum_n \lambda_n f^*_n(\omega'_{\rm in})f_n(\omega_{\rm in})
\end{equation}
Figures~\ref{fig6}(a) and Figures~\ref{fig6}(b) are estimated phase profile $Arg[W_e(\omega'_{\rm in},\omega_{\rm in})]$ and amplitude profile $|W_e(\omega'_{\rm in},\omega_{\rm in})|$ respectively, obtained from measured $Q(\xi,t)$ (see Fig.~\ref{fig4}) using the MLE. The results in Fig.~\ref{fig5} are the estimated complex spectral amplitudes of different input pulses obtained from $W(\omega'_{\rm in},\omega_{\rm in})$ by choosing the $f_n(\omega_{\rm in})$ with highest $\lambda_n$.

In order quantity the quality of our reconstructed TF state, we evalute the fidelity between our estimated and theoretical two-point spectral correlation function. Suppose $W_t(\omega'_{\rm in},\omega_{\rm in})$ and $W_e(\omega'_{\rm in},\omega_{\rm in})$ are the theoretical and estimated quantities. We first write the function $P(\omega'_{\rm i},\omega_{\rm i})={W_t(\omega'_{\rm in},\omega_{\rm in})}W_e(\omega'_{\rm in},\omega_{\rm in})$ and define the fidelity as  
\begin{equation}
	{F} = \int P(\omega_{\rm in},\omega_{\rm in})d\omega_{\rm in} 
\end{equation}
The evaluated fidelity is shown in Fig.~\ref{fig6}.
\section{Definition of similarity S}\label{appen3}
In order to quantify how closely our experimentally measured quantity matches with the corresponding theoretical predictions, we define the similarity $S$ in the following manner. If $t_x$ is the experimentally measured profile and $e_x$ is the corresponding theoretical profile, then the similarity $S$ between experimental and theoretical profile is defined as 
\begin{equation}
	S = \frac{\sum_x e_x t_x}{\sqrt{\sum_x |e_x|^2 \sum_x |t_x|^2}}
\end{equation}
The Similarity $S$ lies between 0 and 1. $S=1$ implies full agreement and $S=0$ implies no agreement. We use this definition to evaluate $S$ for measured $Q-$functions with their corresponding theoretical predictions.
%

\end{document}